\documentclass{elsarticle}

\usepackage{times}
\usepackage{color}
\usepackage{mathtools}
\usepackage{mathbbol}
\usepackage{comment}

\input xy
\xyoption{all}

\usepackage{amsmath}
\usepackage{latexsym,amssymb}
\usepackage{graphics,epsfig}

\newdefinition{definition}{Definition}

\newtheorem{lemma}{Lemma}
\newproof{proof}{Proof}

\DeclareMathSymbol{\fcmp}{\mathrel}{bbold}{\lq\;}

\input{slotsty.sty}
\newcommand{\atomic}[1]{#1}

\begin{document}

\title{A simple proof of three properties on Simpson's 4-slot Algorithm}

\author[xu]{Xu~Wang}\ead{xu.wang.comp@gmail.com}
\author[qiwen]{Qiwen~Xu}\ead{qwxu@um.edu.mo}

\address[xu]{%
School of Software Engineering \\
East China Normal University, China \footnote{$^a$ Part of this work was done while the first author was affiliated with ECNU and the second author was visiting ECNU.}}

\address[qiwen]{
Department of Computer and Information Science \\
University of Macau, China \\}

%\cortext[cor1]{Corresponding authors}
%\cortext[cor1]{Part of this work was done while the first author is affiliated with ECNU and the second author is visiting ECNU.}

\begin{abstract}
In this paper we present an invariance proof of three properties on Simpson's 4-slot algorithm, i.e. \emph{data-race freedom}, \emph{data coherence} and \emph{data freshness}, which together implies linearisability of the algorithm. %and \emph{data freshness}. 
It is an extension of previous works whose proof focuses mostly on data-race freedom. In addition, our proof uses simply inductive invariants and transition invariants~\cite{Andre}, whereas previous work uses more sophisticated machinery like separation logics, rely-guarantee or ownership transfer.

%In comparison with previous works in the literatures, our proof extends 

%Of the three, the proofs of data-race freedom and data coherence are non-trivial. In comparison with previous works in the literatures, which focus mostly on the proof of data-race freedom and use the machinery like separation logics, rely-guarantee or ownership transfer, our proof uses only invariants and transition invariants.
\end{abstract}

\maketitle

\section{Introduction}

In this paper we are going to give a new proof of properties on Simpson's four-slot algorithm~\cite{Simpson90}. The proof consists of two parts: one for the property of \emph{data-race freedom} and the other for the properties of \emph{data coherence} and \emph{data freshness}\footnote{In this paper we adopt a different definition of data coherence than the original one as given in ~\cite{Simpson90} and used in ~\cite{HP2002,JR02,RH09,JP09,Wang}. The original one is essentially data-race freedom in our new setting.}.
%\footnote{There are at least two versions of four slot algorithms by H.R. Simpson~\cite{Simpson90,Simpson97}. The second one uses a weaker atomicity criteria than linearisability.}.
In conjunction, they implies the linearisabilty~\cite{Wing} of the Simpson's asynchronous communication mechanism between a reader and a writer~\cite{PT94}; that is, behaviourally and abstractly the mechanism acts as if it is \emph{a single atomic register} (with linearisabilty semantics) shared between the two parties. The work is inspired by and builds on previous works tackling the same problem by others using a variety of different techniques~\cite{HP2002,JR02,RH09,JP09,Wang}.

%Two important properties are guaranteed by Simpson's asynchronous communication mechanism.
%The first property is \emph{data coherence}, which means that the writer and the reader never access the same slot at the same time. Therefore, the reader always gets a complete value rather than some partially overwritten ones.

In a nutshell, data-race freedom means that \emph{concurrent read and write operations} on the set of (non-atomic) data variables of asynchronous communication mechanism will not race on any member of the set; \emph{race} on a data variable means there is a state on which there exists two threads accessing the same variable simultaneously and incompatibly (e.g. read-write or write-write). The data variables in question consists of a 2x2 array of buffer variables with non-atomic semantics (i.e. the four slots) allocated for storing the contents of the communication. It is due to the \emph{wait-freedom} requirement of the algorithm that \emph{four copies} (of the slot variables) are used in order to implement one copy of the abstract shared register. They are in contrast to the remaining variables of the algorithm which are the (atomic) control ones being put in place to coordinate and synchronise concurrent accesses to data variables for the sake of data-race avoidance.

Data freshness, on the other hand, means that if the duration of a write operation $A$ (also called the round of $A$ operation in the sequel) \emph{precedes} the duration of a read operation $B$, then the contents read by $B$ should be at least as fresh as that written by $A$. Note that
%(to make linearisability valid after its conjunction with data coherence)
the data freshness of all the write operations are \emph{linearly ordered} in the case of the four slot algorithm since it has only one writer; and similarly all the read operations are linearly ordered due to its use of only one reader.

Finally, data coherence means that the order of data writings by the writer should be consistent with the order of data viewing (i.e. readings) by the reader. That is, given a pair of read operations $R1$ and $R2$ reading from the write operations $W1$ and $W2$ resp., if $R1$ is (linearly) \emph{ordered before} $R2$, then $W2$ cannot be possibly (linearly) ordered before $W1$.

\section{Simpson's 4-slot Algorithm}

Now we borrow some exposition from~\cite{Wang} to explain the Simpson's four-slot algorithm.

Simpson's algorithm~\cite{Simpson90}, as shown in Figure~\ref{fig:4slot},
uses four control bits and two pairs of slots as \emph{shared variables} to achieve
asynchronous communication between two threads. In addition to the shared variables, the two threads also use two pairs of local variables, e.g. the pair $wp$ and $wi$ for the writer. %However, the operations on them are invisible to the other thread and can thus coalesce with adjacent operations in the control flow without affecting atomicity.

The reader thread is an (imaginary) loop that repeatedly calls procedure $\pread()$ while the writer thread repeatedly calls procedure $\pwrite(w)$. Commands of the form $\atomic{c}$ in the code of $\pread()$ and $\pwrite(w)$ are atomic commands in which $c$ will be executed in one indivisible step. The four control bits are assumed to be atomic registers %(with sequential consistency semantics)
; thus statements like writing and reading of control bits are atomic commands. The two pairs of slots are assumed to be non-atomic registers, and their assignment and reading are not atomic command.

The cleverness of Simpson's algorithm lies in that the reader and writer can coordinate, via the four atomic control bits, to channel simultaneous requests on the slots to different copies. Thus the accesses to one slot will look as if serial and non-atomic registers will suffice to implement the slots.

\begin{figure}[h]
$
\begin{array}{rl|rl}
\multicolumn{4}{l}{\pshared\ \pslot[2][2] = ((v_{0,0}, v_{0,1}),(v_{1,0}, v_{1,1})), \pindex[2] = (0, 0),
  \platest = 1, \preading = 0 \ \pin\ }\\ \\
  \plocal \ \pwpair=1, &  \pwindex=0;  &
  \ \ \ \ \ \ \ \plocal \ \prpair=0, & \prindex=0, y= \bot; \\
   \quad \ \ &\pwrite(w) = \ \  & \ \ &\pread() =  \\
   \quad {\bf a-2:} \ \ & \atomic{\pwpair := 1-\preading}; \quad \quad &
   \quad {\bf b-3:} \ \ & \atomic{\prpair := \platest};\\
   \quad  {\bf a-1:} \ \ & \atomic{\pwindex := 1 - \pindex[\pwpair]}; \quad\quad &
   \quad {\bf b-2:} \ \ & \atomic{\preading := \prpair};\\
   \quad  {\bf a:} \ \  & \pslot[\pwpair][\pwindex] := w; \quad\quad &
   \quad {\bf b-1:} \ \ & \atomic{\prindex := \pindex[\prpair]};\\
   \quad  {\bf a+1:} \ \ & \atomic{\pindex[\pwpair] := \pwindex}; \quad\quad &
   \quad {\bf b:} \ \ & y := \pslot[\prpair][\prindex];\\
   \quad  {\bf a+2:} \ \ & \atomic{\platest := \pwpair}; \quad\quad &
   \quad  {\bf b+1:} \ \ & \preturn\ y;\\
\end{array}
$
\caption{The four-slot algorithm}
\label{fig:4slot}
\end{figure}

In $\pwrite(w)$ the local variables $wp$ and $wi$ act as pointers pointing to resp. a pair and a slot in the pair. Collectively they identify the slot the writer is going to write to. The values of these pointers depends on the values of control bits $\preading$ and $\pindex$. $\preading$ is a pointer used by the reader to publish the pair it is going to read from, while $\pindex$ are two pointers pointing to resp. the slots holding the freshest value in each pair. The strategy of the writer, upon each invocation of $\pwrite(w)$, is to move away from the pair the reader is working on and select the slot not holding the freshest value to write to. After writing to the slot, the writer updates the relevant pointer in $\pindex$ (to point to the new freshest) and publish its latest location (i.e. the pair it just worked on) in $\platest$.

Similarly, in $\pread()$ $rp$ and $ri$ point to the slot the reader is going to read from. The strategy of the reader is to track the latest location of the writer (by reading $\platest$) and read the freshest value in the location (as pointed to by $\pindex$). However, notice that the reader
updates $\preading$ before the read starts, while the writer
updates $\platest$ and $\pindex$ after the write is finished. The order in which
the control variables are updated in each thread
is very crucial for achieving asynchronous communication.

Data-race freedom is mainly achieved in the writer's strategy,
%. The second property is \emph{data freshness}, whilst D
data freshness is mainly achieved in the reader's strategy; and data coherence is achieved by the collaboration of the two.

\noindent
{\bf Semantics of non-atomic variable access.} In this paper, instead of using a non-atomic semantic model that interprets the execution of a non-atomic action as an interval (e.g. a pair of transitions), we will use atomic interleaving model to give semantics to the four-slot algorithm, where we assume all action (i.e. command) executions are atomic. The rationale here is that the latter is faithful to the former on all execution sequences up to the first non-atomic data access. 

If furthermore the data-race freedom can be proved in the atomic model of the four-slot algorithm, we can show the two models coincide. Our argument is as follows. 

We say a state $s$ is \emph{as-if atomic} if for all data variables %either 
there is at most one thread in that state that is accessing the data variable. % or all threads accessing it are read access. 
Then, for all execution sequences of the program in the atomic model, the first state encountered that is enabled with a non-atomic action in Simpson's algorithm will be as-if atmoic (due to the data-race freedom and 1-reader and 1-writer nature of the algorithm). All non-atomic accesses in an as-if atomic state can be treated as atomic; and then inductively it can be shown that the second and all subsequent non-atomic accesses can be treated as atomic. So data-race freedom with atomicity assumption implies data-race freedom without such assumption, and the two models coincide.

\section{Semantic basis of induction and deduction rules}

%\noindent
%{\bf Notation on binary relations.} Given a set $S$, we use $R$, $R'$, $R_1$, $R_2$, etc. for a \emph{binary relation} over $S$, $^{\bullet}\!R$ and $R ^{\bullet}$ for the \emph{domain} and \emph{range} of $R$, $D^{=}$ and $D^{\times}$ for the \emph{identity} and \emph{universal} relations over \emph{the restricted domain} $D \subseteq S$, %and for the with domain restricted to $S_x \subseteq S$,
%$R^{-1}$ and $\overline{R}$ for the \emph{inverse} and \emph{complement} (w.r.t. $S^{\times}$) relations
% of $R$, % when $T$ satisfies $[T) = (T]$.
% $R_1 \circ R_2$ for \emph{relation composition} of $R_1$ and $R_2$ (being an extension of function composition), and $R(D)$ for the \emph{image} of $R$ when its domain being restricted to $D \subseteq S$.

%under the side condition
%under the restriction
%that $dom(T) = ran(T)$.
% $[T) \times (T]$
%the \emph{universal relation} $U$ (i.e. $S \times S$),
%the \emph{domain-restricted } ,

\noindent
{\bf Transition system.} Given a set of states $S$, we can built a \emph{state-transition system} $T\!S= (S, S_0, \Delta)$ s.t. $S_0 \subseteq S$ is the set of \emph{initial states} and $\Delta$ is a set of \emph{small-step transitions} (aka indivisible transitions).
%{\bf Transition system.} Given a set of global actions $\mathcal{A} = \{A_i | 1\leq i \leq n\}$ over the state space $S$, we can built a (global-state) \emph{transition system} $T\!S= (S, S_0, \Delta)$ s.t. $S_0 \subseteq S$ is the set of \emph{initial states} and $\Delta = \bigcup_{A_i \in \mathcal{A}} A_i$ is a set of \emph{small-step transitions} (aka atomic transition).

%We can also build an \emph{abstract transition system} as a footprint-labelled transition system $T\!S^{fp}= (S, S^0, FP, \Delta^{fp})$ s.t. $FP \subseteq \mathcal{FP}$ is a set of factorised footprints and $\Delta^{fp} \subseteq S \times FP \times S$ is a set of footprint-labelled transitions and $(s, \langle fp \rangle, s') \in \Delta^{fp}$ for some $fp \in FP$ iff there exists some action $A_i \in \mathcal{A}$ s.t. $(s,s') \in A_i$ and $A_i$ is a global action with factorised side effect $fp$.

%A (state) \emph{transition system} $T\!S= (S, S^0, \Delta)$ consists of a set of \emph{states} $S$, a distinguished subset $S^0 \subseteq S$ of \emph{initial states}, and a set of \emph{small-step transitions} (aka atomic transition) $\Delta \!:\! S \times S$.

For $T\!S$, we use $S_0^R$ to denote the set of \emph{reachable states} in $T\!S$ \emph{from $S_0$}, $\Delta^R$ to denote the set of \emph{reachable transitions} in $T\!S$, and $\Delta^+$ to denote the transitive closure of $\Delta$; and we call $(s,s') \in \Delta^+$ a \emph{big-step transition} (aka divisible transition) of $T\!S$ since $s$ needs to traverse a \emph{non-trivial chain} of small-step transitions to reach $s'$.

\subsection{State invariants and transition invariants}

%An \emph{abstract transition system} is a footprint-labelled transition system.

\noindent
{\bf Predicate and program.}
We assume a state $s \in S$ denotes a \emph{valuation} over a set $X$ of variables, where $X'$ is the set of primed counterparts of $X$. We use $s(x)$ to denote the value held by the variable $x \in X$ at state $s$. A \emph{state predicate} $p$ (over $X$) denotes a subset $[p]$ of $S$. A \emph{state-pair predicate} $pp$ (over $X$ and $X'$) denotes a subset $[pp]$ of $S \times S$.

We use $p[X'/X]$ to denote the substitution of $X$ variables occurring in $p$ by their primed counterparts from $X'$. A command $c$ in a program $PG$ can be written as a state-pair predicate whilst the initialisation $init$ of $PG$ can be written as a state predicate.

%$(\delta_X, pp_e)$ s.t. $\delta_X \subseteq X$ specifies the set of variables $c$ has ownership (i.e. can modify)
% whilst $pp_e$ specifies the side effect (i.e. sub-state modification) produced by command $c$ on the global state. %If we define $T(X \setminus \delta)$ to be the set of transitions s.t.
%$\langle c \rangle$ denotes the set of transitions made by command $c$ and is defined to be $[pp_e] \cap \mbox{\boldmath $=$}_{\lfloor X \setminus \delta \rfloor}$, where $(s,s') \in \mbox{\boldmath $=$}_{\lfloor X \setminus \delta \rfloor}$ iff $s[x] =s'[x]$ for all $x \in X \setminus \delta$. Alternatively, $c$ can also be directly equated to a state-pair predicate $pp_c \triangleq pp_e \wedge \textsc{eq}_{X \setminus \delta}$ where $\textsc{eq}_{X_0} \triangleq \bigwedge_{x \in X_0} x=x'$.

\noindent
{\bf Invariant.} A state predicate $p$ is an \emph{invariant} (for $T\!S$) iff the set of states it denotes, i.e. $[p]$, is a superset of $S^R$; and furthermore it is an \emph{inductive invariant} iff $S_0 \subseteq [p]$ and $\Delta([p]) \subseteq [p]$, where $\Delta([p])$ is the image produced by the relation $\Delta$ when its domain is restricted to $[p]$.

\noindent
{\bf Transition invariant.} %A state-pair predicate $pp$ is an \emph{atomic transition invariant}, i.e. invariant for small-step transitions in $T\!S$, iff the set of state pairs it denotes, i.e. $[pp]$, is a superset of $\Delta^R = \Delta \cap S_0^R \times S$; and $pp$ is an \emph{inductive atomic-transition invariant} iff $\Delta \subseteq [pp] $.
A state-pair predicate $pp$ is a (global) \emph{transition invariant}\cite{Andre}, i.e. invariant for big-step transitions %(spanning arbitrary non-zero length)
in $T\!S$, iff the set of state pairs it denotes, i.e. $[pp]$, is a superset of $(\Delta^R)^+$ (i.e. the set of \emph{reachable big-step transitions}); and $pp$ is an \emph{inductive transition invariant} iff $\Delta \cap (S_0 \times S) \subseteq [pp]$ and $[pp] \circ \Delta \subseteq [pp]$, where $[pp] \circ \Delta$ stands for \emph{relation composition} of $[pp]$ and $\Delta$ (being an extension of function composition).

%$\Delta^R = \Delta \cap S_0^R \times S$
%transitive closure of $\delta^r = \delta \cap s_0^r \times s$, i.e.

\subsection{Induction and deduction rules for invariance reasoning}

The invariance principle of assertional reasoning for concurrent programs lies in the use of invariants of various forms (e.g. state- or transition- invariants) to express everything, from properties and specifications to actions and programs. The reasoning consists of decomposing complex invariants into simple ones and finding inductive invariants from which these simple invariants can be deduced. Below we use $Inv(PG)$ to denote the set of all state- or transition- invariants for program $PG$, and use $inv \in Inv(PG)$ to mean the (state- or state-pair-) predicate $inv$ is a state- or transition- invariant.
%  decompose

An invariant of a concurrent program is a condition that holds true on all reachable global state of the program. The most effective way to establish an invariant is by induction on the initialisation $init$ and all commands of the program, which gives rise to the so-called \emph{inductive invariants}.

\[ \mfrac{init  \!\implies\! p \ \ \ \ \forall \, c \in \textit{PG} : p \wedge c \implies p\texttt{[X'/X]}}{p \in \textit{Inv(PG)}} \tag{\textsc{Induction-S}}\]
%\[ \mfrac{init  \!\implies\! inv \ \wedge \ \forall \, c \in \textit{PG} : \{inv\} \ c \ \{inv\}}{inv \ \, \textsl{is \ an \ invariant}} \tag{\textsc{Induction}}\]

\noindent
Similarly we have \emph{inductive transition invariants}:

\[ \mfrac{\forall \, c \in \textit{PG} : (pp\texttt{[X$^{\circ}\!\!$/X'\!\!]} \wedge c\texttt{[X$^{\circ}\!\!$/X]} \!\implies\! pp)
\ \wedge  \exists p \in \textit{Inv(PG)} : (p \wedge c \!\implies\! pp)}
{pp \in \textit{Inv(PG)}} \tag{\textsc{Induction-T}}\]

%\[ \mfrac{init \wedge \bigvee \textit{PG} \!\implies\! pp \ \ \ \   \forall \, c \in \textit{PG} : pp\texttt{[X$^{\circ}\!\!$/X'\!\!]} \wedge c\texttt{[X$^{\circ}\!\!$/X]} \!\implies\! pp}{pp \in \textit{Inv(PG)}} \tag{\textsc{Induction-T}}\]

%\[ \mfrac{ \forall \, c \in \textit{PG} : init \wedge  c \!\implies\! pp \wedge pp\texttt{[X$^{\circ}\!\!$/X'\!\!]} \wedge c\texttt{[X$^{\circ}\!\!$/X]} \!\implies\! pp}{pp \ \, \textsl{is a transition invariant}} \tag{\textsc{Induction-T}}\]

In order to simplify our proof, we introduce a new technique we call \emph{inductive subject to}. For instance, given a supporting set of invariants $I$, we say a state predicate $p$ is an \emph{inductive invariant subject to} $I$, iff it can be established inductively by checking that, for all $c_i$ in $PG$, there exists some (state- or transition-) invariant $inv_i \in I$ s.t.

\[ p \wedge inv_i \wedge c_i \implies p\texttt{[X'/X]} \]

\noindent
holds. The same is true for the inductive subject-to technique for transition invariants.

\[ \mfrac{init  \!\implies\! p \ \ \ \  \forall \, c_i \in \textit{PG}: \exists inv_i \in \textit{Inv(PG)} : p \wedge (inv_i \wedge c) \implies p\texttt{[X'/X]}}{p \in \textit{Inv(PG)}} \tag{\textsc{Subject-S}}\]

\[ \mfrac{\parbox{4.0in}{
$\forall \, c_i \in \textit{PG}: \exists inv_i \in \textit{Inv(PG)} : (pp\texttt{[X$^{\circ}\!\!$/X'\!\!]} \wedge (inv_i \wedge c_i) \texttt{[X$^{\circ}\!\!$/X]} \!\implies\! pp)$
\\
${\ \ \ \ \ \ \ \ \ \ \ \ \ \ \ \ \ \ \ \ \ \ \ \  } \ \wedge \ (\exists p_i \in \textit{Inv(PG)} : p_i \wedge (inv_i \wedge c_i) \!\implies\! pp)$
}
}{pp \in \textit{Inv(PG)}} \tag{\textsc{Subject-T}}\]

%\[ \mfrac{\parbox{4.0in}{
%{\ \ \ \ \ \ \ \ \ \ \ \ \ \ \ \ \ \ \ \ \ \ \ \ \ \ \ \ \ \ \ \ \ \ \ \ \ \ \ \  }
%$init \wedge \bigvee \textit{PG} \!\implies\! pp$
%\\
%$\forall \, c_i \in \textit{PG}: \exists inv_i \in \textit{Inv(PG)} : pp\texttt{[X$^{\circ}\!\!$/X'\!\!]} \wedge (inv_i \wedge c_i) \texttt{[X$^{\circ}\!\!$/X]} \!\implies\! pp$
%}
%}{pp \in \textit{Inv(PG)}} \tag{\textsc{Subject-T}}\]

%\[ \mfrac{inv \in \textit{Inv(PG)} \wedge \forall \, c \in \textit{PG} : pp\texttt{[X$^{\circ}\!\!$/X'\!\!]} \wedge (inv \wedge p \wedge c) \texttt{[X$^{\circ}\!\!$/X]} \!\implies\! pp}{p : pp \in \textit{lt-Inv(PG)}} \tag{\textsc{L-induction-T}}\]

%\[ \mfrac{init  \!\implies\! p \wedge \forall \, A \in \textit{PG} : \{p\} \ A \ \{p\}}{p \ \, \textsl{is an invariant}} \tag{\textsc{Induction-S}}\]

%\[ \mfrac{inv \ \textsl{is \ an \ invariant} }{inv \wedge inv' \ \textsl{is \ an \ invariant}} \tag{\textsc{Induction-S}}\]

After the establishment of all the inductive invariants, we often use implication and conjunction to obtain new state- and transition- invariants, which are not necessarily inductive.

%\[ \mfrac{ \{p \wedge \textrm{pre}(pp)\} \  c \ \{qq\} \wedge \ qq_{(X,X^{\circ})} \wedge \textrm{pre}(pp)_{X^{\circ}} \wedge \textsc{spec}(c)_{(X^{\circ},X')} \implies qq}{qq@p\!:\!pp} \tag{\textsc{Induction-L}}\]

\[ \mfrac{pp, \ pp' \in \textit{Inv(PG)} \ \ \ \ \ \ 
pp\texttt{[X$^{\circ}\!\!$/X'\!\!]} \wedge pp'\!\texttt{[X$^{\circ}\!\!$/X\!]} \!\!\implies\!\! pp'' 
}{pp'' \in \textit{Inv(PG)}}
\tag{\textsc{Composition}}\]

\[ \mfrac{inv' \!\implies\! inv \ \ \ inv' \in \textit{Inv(PG)}}{inv \in \textit{Inv(PG)}}
% \ \ \ \mfrac{pp' \!\implies\! pp \ \ \ pp' \in \textit{t-Inv(PG)}}{pp \in \textit{t-Inv(PG)}}
\tag{\textsc{Consequence}}\]

\[ \mfrac{inv, \ inv' \in \textit{Inv(PG)}}{inv \wedge inv' \in \textit{Inv(PG)}}
%\ \ \ \mfrac{pp, \ pp' \in \textit{t-Inv(PG)}}{pp \wedge pp' \in \textit{t-Inv(PG)}}
\tag{\textsc{Conjunction}}\]

\section{An inductive invariant proof of data-race freedom}

In this section, we present our simple proof of data-race freedom, which is based on the invariance principle of assertional reasoning for concurrent programs\footnote{The proof was first found by the second author in~\cite{Xu09}.}. That is, to establish an invariant, we decompose complex invariants into simple ones and then find inductive invariants to which these simple invariants are consequences. %  decompose

% lies in the use of invariants of various forms (e.g. state or transition invariants) to express everything, from properties and specifications to actions and programs.
%An invariant of a concurrent program is a condition that holds true on all reachable global state of the program.
%The most effective way to establish an invariant is by induction on all commands (including the initialisation, i.e. $init$) of the program, which gives rise to the so-called \emph{inductive invariants}. Hence, the invariance reasoning can be reduced to 1) decomposing complex invariants into simple ones and 2) finding inductive invariants to which these simple invariants are consequences. %  decompose

We can formalise the property of data-race freedom as follows:

\[\alpha = a \wedge \beta= b \!\implies\! (wp \neq rp \vee wi \neq ri) \tag{\textsc{Race-freedom}}\]

\noindent
where $\alpha$ and $\beta$ are program counters respectively for the writer and for the reader.

The invariant says that if there is a global state (i.e. a value assignment to all the local and shared variables of the programs including program counters) in which the reader is reading a slot and the writer is writing a slot, then the two slots must be different, i.e. $wp \neq rp \vee wi \neq ri$, which is essentially the freedom of data races on the slots. \textsc{Race-freedom} can be further strengthened to \textsc{Race-freedom-ex}:

\vspace{-3mm}

\[\alpha \in \{a, a+1 \} \wedge\, \beta \notin \{b-2, b-1\} \!\implies\! (wp \neq rp \,\vee\, wi \neq ri) \tag{\textsc{Race-freedom-ex}}\]

\noindent
%\vspace{3mm}
which is easier to prove by decomposition into three conditions:

\[ \alpha \in \{a, a+1 \} \!\implies\! wi \neq li[wp] \tag{\textsc{Cond1}}\]

\vspace{-3mm}

\[ \beta \notin \{b-2\}  \!\implies\! r= rp \tag{\textsc{Cond2}}\]

\vspace{-5mm}

\[ \alpha \in \{a-1, a, a+1 \} \wedge \beta \notin \{b-2, b-1\} \!\implies\! (wp \neq r \lor ri = li[rp]) \tag{\textsc{Cond3}}\]

%\vspace{3mm}
\noindent
where it can be deduced that $\textsc{Cond1} \wedge \textsc{Cond2} \wedge \textsc{Cond3} \!\implies\! \textsc{Race-freedom-ex}$.

\textsc{Cond1} is an inductive invariant, which can be established by checking initialisation as well as all command $c$ in the four-slot program. Actually, the check is trivial for all commands except for $a-1: \  \atomic{\pwindex := 1 - \pindex[\pwpair]}$ which update $\alpha$ from $a-1$ to $a$ since for the other commands either the antecedents of the implications remain invalid or the commands do not update the variables $\pwindex$ and $\pindex[\pwpair]$.

\textsc{Cond2} is an inductive invariant 
whose check holds trivially for initialisation and all commands except for $b-2: \ \atomic{\preading := \prpair}$ and $b-1: \ \atomic{\prindex := \pindex[\prpair]}$. 
With the use of rule \textsc{Consequence}, we know \textsc{Cond2} is an invariant.

%but is not inductive. A strengthening of \textsc{Cond2}

%\[\beta \notin \{b-3, b-2\} \!\implies\! r= rp \tag{\textsc{Cond2}\texttt{'}}\]

%\noindent
%however, is an inductive invariant, 

Similarly, \textsc{Cond3} is an inductive invariant%but not inductive; one of its strengthening as follows is an inductive invariant:
%\[\alpha \in \{a-1,a, a+1, a+2\} \wedge \beta \neq b-1 \!\implies\! (wp \neq r \lor ri = li[rp]) \tag{\textsc{Cond3}\texttt{'}}\]
%\noindent
, whose check holds trivially for initialisation and all commands except for\,:

\[
\begin{array}{rl}
a-2: & \atomic{\pwpair := 1-\preading} \\
a-1: & \atomic{\pwindex := 1 - \pindex[\pwpair]} \\
b-1: & \atomic{\prindex := \pindex[\prpair]}
\end{array}
\]

%is an inductive invariant. Readers are encouraged to verify this by going through each command $cmd$ of the program such that \{\textsc{Cond1}\} \ cmd \ \{\textsc{Cond1}\} forms a Hoare logic triple.

\section{An inductive proof of data coherence and data freshness}

Our proof of data 
freshness and 
coherence demands the decoration of the original program with auxiliary variables. In this paper we add timestamps to the original program. That is, $wtm$ is a counter used to timestamp each round of the write operation by the writer so that all writes of shared variables in the same round are decorated with the same timestamp.
%Each shared variable written by the writer are therefore decorated with a timestamp.
For instance, variable $li[x]$ becomes variable $LI[x]$, which can be understood as a record datatype consisting of two fields: $LI[x].val$ and $LI[x].tm$; the former holds the value (say $v$) originally held by $li[x]$ while the latter holds the timestamp marking the exact round at which $v$ is written into $li[x]$. Similarly, $D[x][y]$ are decorated version of $d[x][y]$.

%For instance, variable $l$ becomes variable $L$, which can be understood as a record datatype consisting of two fields: $L.val$ and $L.tm$; the former holds the value (say $v$) originally held by $l$ while the latter holds the timestamp pinpointing the exact round at which $v$ is written into $l$. Similarly, $LI[\ ]$ and $D[\ ][\ ]$ are decorated versions of $li[\ ]$ and $d[\ ][\ ]$.

\begin{figure}[h]
$
\begin{array}{rl|rl}
\multicolumn{4}{l}{\pshared\ D[2][2] = (((v_{00},0), (\bot,\bot)), ((v_{10},1), (\bot,\bot))), LI[2] = ((0,0), (0,1)),} \\
\multicolumn{4}{l}{\ \ \ \ \ \ \ \ \ \ \ \ \ \ l = 1, \preading = 0, \ \textcolor{blue}{wtm,rtm=1,0} \ \ \ \pin \ }\\ \\
  \plocal \ \pwpair=1, & \!\!\! \pwindex=0;  &
  \plocal \ \prpair=0, & \!\!\! \prindex=0, y= \bot; \\
    &\!\!\! \pwrite(w) = &  &\!\!\! \pread() =  \\
   {\bf a-2:} &\!\!\! \atomic{\textcolor{blue}{wtm++;}\ \pwpair := 1-\preading};  &
   {\bf b-3:} &\!\!\! \atomic{ \prpair := l};\\
   {\bf a-1:} &\!\!\! \atomic{\pwindex := 1 - LI[\pwpair].val}; &
   {\bf b-2:} &\!\!\! \atomic{\preading := \prpair};\\
%   \quad {\bf b-2:} \ \ & \atomic{{\bf If\ } \preading \neq \prpair {\ \bf Then\ } \{li[r]:=\bot;\preading := \prpair\}};\\
   {\bf a:} &\!\!\! D[\pwpair][\pwindex] := (w, wtm); &
   {\bf b-1:} &\!\!\! \atomic{(\prindex,\textcolor{blue}{rtm}) := LI[\prpair]};\\
   {\bf a+1:} &\!\!\! \atomic{LI[\pwpair] := (\pwindex, wtm)}; &
   {\bf b:} &\!\!\! (y,\textcolor{blue}{rtm}) := D[\prpair][\prindex];\\
   {\bf a+2:} &\!\!\! \atomic{l := \pwpair}; &
   {\bf b+1:} &\!\!\! \preturn\ y;\\
\end{array}
$
\caption{The four-slot algorithm with timestamps}
\label{fig:4slot-mod}
\end{figure}

All the write operations (defined by the procedure $\pwrite(w)$) are linearly ordered and it is the same for all the read operations (i.e. the procedure $\pread()$), giving rise to a pair of linear orders. Within one linear order, we have relations like (linearly) \emph{ordered after} and \emph{immediately ordered after}.

Between linear orders, we say an operation $A$ from one linear order $O$ \emph{precedes} an operation $B$ from another linear order $O'$ iff in an interleaved %(i.e. sequential consistent) 
execution of the concurrent programs, the last command in $A$ are executed before the first command in $B$; $A$ \emph{overlaps} $B$ iff neither $A$ precedes $B$ nor $B$ precedes $A$; and $A$ \emph{immediately precedes} $B$ iff $A$ precedes $B$ and there is no other operation $A'$ ordered after $A$ in $O$ such that $A'$ precedes $B$.

%\begin{proof}
%$R$ precedes $W$ implies $R$ cannot read from $W$: $wtm \geq LI[rp].tm$ is an invariant from which we can derive $wtm \neq wtm' \implies LI[rp].tm < wtm'$;
%$W$ immediately precedes $R$ implies $R$ cannot read from $W^-$ ordered before $W$: $\beta \in \{b-3..b+1\} : wtm > t \wedge wtm' > t \implies \beta \in \{b\} : rtm' > t$.

%the invariant $wtm \geq LI[rp].tm$ induces $wtm \neq wtm' \implies wtm < LI'[rp'].tm$.

%is an invariant from which we can derive $wtm \neq wtm' \implies LI[rp].tm < wtm'$

%Since from the above lemma we have $W$ immediately precedes $R$ iff at $b-3$, $rp := l \in R$ reads from $l := wp \in W$. Furthermore we have $rp := l \in R$ is ordered before $(y,\textcolor{blue}{rtm}) := D[\prpair][\prindex] \in R$ in the executions. Use the monotonicity lemma above we have that the value held by variable $rtm$ at $b-3$ in $R$ cannot be larger than the value held by $rtm$ at $b+1$.
%\end{proof}

\subsection{The proof of data coherence}

In this subsection we present a series of state- and transition- invariants for the Simpson's four slots program in order to prove the main lemma of this paper.

%and $\alpha \neq a+1 \wedge p= wp.val \implies li[p].tm = d[p][li[p].val].tm$ are invariants.
%the write operations are linearly ordered with strictly increasing timestamps 2) within one operation the commands will update (i.e. increase the value of) the three timestamps above in the reverse order of the inequality.

%$LI[r].tm \leq LI[L.val].tm$ is an invariant.

\begin{lemma}[Location monotonicity]
For all timestamped variable $x \in \{LI[p], D[p][i] \,$ $| \,p, i \in \{0,1\}\}$, the transition invariant below holds:
\[x.tm \leq x'.tm \]

%For any execution $\theta$ (sequence of states) of the above timestamped program, any pair of global states $s$ and $s'$ in $\theta$ and any timestamped variable $x \in \{L, LI[p], D[p][i] \, | \,p, i \in \{0,1\}\}$, if $s$ precedes $s'$ in $\theta$, then we have $s[x.tm] \leq s'[x.tm]$.

%the valuation of variable $x$ in $s$ and $s'$, denoted $s[x]$,
\end{lemma}
\begin{proof}
$x.tm \leq x'.tm$ is an inductive transition invariant subject to the inductive invariant $\forall x \in \{LI[p], D[p][i] \,| \,p, i \in \{0,1\}\}: x.tm \leq wtm$.
%Use the rule \textsc{Induction-T} to derive $x.tm \leq wtm \wedge x.tm \leq x'.tm \wedge x'.tm \leq wtm'$ as an inductive transition invariant.
%Initially $x=x'=x_0$ it holds; for every small step transitions in $T$ it holds; for all the small step transitions in $T$ it is robust when composed either as pre-action or post-action.

%To prove $p \wedge p': pp$ is an inductive transition invariant, we prove $init \!\implies\! p$
%Obvious since all the writes on these variables are linearly ordered with respect to which the timestamp counter is strictly increasing.
\end{proof}

\begin{lemma}[Reader monotonicity]
%For all timestamped variable $x \in \{LI[p], D[p][i] \,$ $| \,p, i \in \{0,1\}\}$, the transition invariant below holds:
%\[x.tm \leq x'.tm \]
The transition invariant below holds:
\[rtm \leq rtm'\]
\end{lemma}
\begin{proof}
%From the a) and 3) above we can show that
$rtm \leq rtm'$ is an inductive transition invariant subject to the invariants:

\[ \beta = b \!\implies\! rtm = D[rp][ri].tm \tag{\textsc{CondA}}\]

%\vspace{-4mm}

%\[ \beta \notin \{b-2, b-1\} \wedge \alpha \in \{a, a+1\} \!\implies\! wp \neq rp \vee wi \neq ri \tag{\textsc{CondA'}}\]

\vspace{-3mm}

\[ \beta \in \{b-2, b-1\}  \!\implies\! rtm \leq LI[rp].tm \tag{\textsc{CondB}}\]

\vspace{2mm}

The \textsc{CondA} is an inductive invariant subject to the invariant a) $\forall p \in \{0,1\}: LI[p].tm = D[p][LI[p].val].tm$, the inductive invariant $\beta \notin \{b-2, b-1\} \!\implies\! ri = li[rp]$ and the \textsc{Race-freedom-ex}. The a) is an inductive invariant subject to the \textsc{COND1} (i.e. $\alpha=a \implies wi \neq LI[wp].val$) and the inductive invariant $\alpha=a+1 \implies wtm=D[wp][wi].tm$.% and a strengthening of \textsc{Race-freedom}, i.e. $\alpha \in \{a, a+1\} \wedge \beta \notin \{b-2,b-1\} \!\implies\! (wp \neq rp \vee wi \neq ri)$.

The \textsc{CondB} is an inductive invariant subject to the location monotonicity and the invariant $\beta = b-3 \!\implies\! rtm \leq LI[rp].tm \leq LI[l].tm$, which is a %an %inductive invariant subject to 1) %and the \textsc{Race-freedom-ex}. %$\beta \neq b-2 \!\implies\! rp = r$
 consequence of the conjunction of the inductive invariant $\beta \notin \{b-2, b-1\} \!\implies\! rtm \leq LI[rp].tm$ (subject to the location monotonicity and the \textsc{CondA}) and the transition invariant 1) $LI[l].tm \leq LI'[l'].tm$.  The 1) is an inductive invariant subject to the location monotonicity and the inductive invariant $\alpha = a+2 \!\!\implies\!\! LI[wp].tm = wtm$ and $\forall i \in \{0,1\}: LI[i].tm \leq wtm$.

\end{proof}

\noindent{\bf Data coherence:}

\begin{quote}
If a read operation is ordered before another read operation, the former cannot read from a slot which is strictly more fresh than the one read by the latter.
\end{quote}
\begin{proof}
Use the reader monotonicity lemma above.
\end{proof}

\subsection{The proof of data freshness}

%\begin{lemma}
%A write operation $W$ immediately precedes a read operation $R$ in an execution iff $rp := l \in R$ reads from $l := wp \in W$.

%$\beta \in \{b-3..b+1\} : wtm > t \wedge wtm' > t \implies \beta \in \{b\} : rtm' > t$.

%$\alpha \neq a + 2 \vee wtm \geq LI[wp].tm \vee $.

%$rp := l \in R$ reads from $l := wp \in W$.
%\end{lemma}
%\begin{proof}
%Obvious since $rp := l$ is the first command in $R$ and $l := wp$ is the last command in $W$.
%\end{proof}

%\begin{lemma}
%A read operation $R$ immediately precedes a write operation $W$ in an execution implies all readings made in $R$ must have strictly smaller timestamps than that of $\, W$'s writings.
%\end{lemma}
%\begin{proof}
%Obvious since the global state read by $R$ is created by the write operations (strictly) linearly ordered before $W$, which have timestamps strictly smaller than that of $W$.
%\end{proof}

\vspace{1mm}
\noindent{\bf Data freshness:}

\begin{quote}
A read operation $R$ can only read from a write operation overlapping $R$ or immediately preceding $R$.
\end{quote}

\begin{proof}
We need to prove two cases: $R$ precedes $W$ implies $R$ cannot read from $W$ and $W$ immediately precedes $R$ implies $R$ cannot read from $W^-$ ordered before $W$. 

The first case is implied by the transition invariant $\beta' = b-2 \wedge \beta = b+1 \implies rtm < wtm'+1$. It is an inductive transition invariant subject to the inductive invariant e) $\forall p,i \in \{0,1\}: D[p][i].tm < wtm +1$, the inductive invariant f) $\beta \in \{ b, b+1 \} \implies rtm = D[ri][rp].tm$ (subject to the \textsc{Race-freedom} and \textsc{CondA}), and the reader monotonicity.

% (derivable by the conjunction rule from %the \textsc{CondA} above, 
%the inductive invariants $\forall p,i \in \{0,1\}: wtm - 1 \leq D[p][i].tm < wtm +1$ and the reader monotonicity). %the inductive transition invariant $wtm \leq wtm'$,
%the lemma above.

%implied by the transition invariant $\beta=b-3 \wedge \beta' \notin \{b-3,b-2,b-1\} \implies rtm' > wtm-2$ (derivable by the conjunction rule from the inductive invariants $\beta \notin \{b-3,b-2,b-1\} \implies rtm \leq LI[rp].tm$,
%$\forall x \in \{LI[p], D[p][i] \,$ $| \,p, i \in \{0,1\}\}:
%$LI[l].tm \geq wtm-1$
%LI[xtm \neq rtm' \implies rtm' \geq wtm-1$
% and the \textsc{CondB}).
%can be induced from
%using the first lemma and the fact that $rp := l$ in $R$ reads from $l := wp$ in $W$ implies

The second case is implied by $\beta=b-3 \wedge \beta' = b+1 \!\!\implies\!\! rtm' \geq wtm-1$.
%$\beta=b-3 \wedge wtm=a \!\!\implies\!\! \beta' \neq b \vee rtm' \geq a-1$. 
It is a consequence of the composition of the reader monotonicity and the invariant k) $\beta=b-3 \wedge \beta' = b \!\!\implies\!\! rtm' \geq wtm-1$.
The k) is a composition of the invariants $LI[l].tm \geq wtm-1$ (which can be established by sequential deduction on the writer's thread only), $\beta=b-3 \wedge \beta' = b-2 \!\!\implies\!\! LI'[rp'].tm \geq LI[l].tm$ (inductive subject to $LI[l].tm \leq LI'[l'].tm$ and location monotonicity), the location monotonicity, and $\beta=b-1 \wedge \beta' = b \!\!\implies\!\! rtm' \geq LI[rp].tm$ (inductive subject to the location monotonicity).

%$\beta=b-1 \wedge \beta' = b \!\!\implies\!\! rtm' \leq LI[r]$ and 

%the inductive transition invariant $wtm \leq wtm'$ and 
The invariant $LI[l].tm \leq LI'[l'].tm$ can be established by sequential deduction on the writer's thread only.

\end{proof}

%\[ \beta  \in \{b+1,b-4,b-3,b-2,b-1\} \implies rtm \leq rtm' \tag{\textsc{Case1}} \]
%\[ \beta  \in \{b\} \implies rtm = rtm' \tag{\textsc{Case2}} \]

%\end{comment}

\section{Discussion}

Our work differs significantly from existing works on four-slot algorithm verification~\cite{HP2002,JR02,RH09,JP09,Wang}. ~\cite{JR02} uses model checking whilst the others, like us, uses theorem proving. ~\cite{JR02} encodes and verifies all three properties directly or indirectly whilst the theorem proving works focus mostly on the verification of the data-race freedom.

%We adopt, like~\cite{JR02,HP2002,JP09}, linearisability and Lamport's atomicity hierarchy (for registers) as our criterion of atomicity. 

% whereas previous work~\cite{} uses mostly %members of 
%Lamport's atomicity hierarchy for registers. 
On the semantic modelling of non-atomic data access, ~\cite{JR02} uses `random' variables with non-deterministic assignment whilst we use more a reductionist strategy to collapse such data access to atomic actions. %, instead of modelling it as `random' variables with non-deterministic assignment like~\cite{JR02}.

Lastly, our proof adopts the global approach of assertional reasoning whilst the previous work~\cite{RH09,Wang} uses more thread-local approach with rely/guarantee, separation logics and ownership transfer.

\section{Conclusion}

We have given a simple proof of data-race freedom, data coherence and freshness on Simpson's four-slot algorithm, which, in conjunction, implies linearisability. %\footnote{The parts for data-race freedom and data coherence are esp. interesting.}
It uses only the inductive state- and transition- invariants for the proof of the three properties, which significantly simplified previous works (mostly focusing on data-race freedom) that uses separation logics, rely/guarantee, ownership transfer or their combinations for the same purpose.

%especially the parts for data-race freedom and data coherence are non-trivial.

%That is unlike
\section*{Acknowledgement}
We benefit from discussion with Dr Mengda He and we thank encouragements from Prof Cliff Jones.

\bibliographystyle{plain}
\bibliography{slotbib}
\end{document}